\documentclass[12pt,preprint]{aastex}

\shorttitle{Unipolar Induction Star-Disk Interaction}
\shortauthors{Tsui}

\begin{document}

\title{Unipolar Induction Star-Disk Interaction}
\author{K.H. Tsui}
\affil{Instituto de F\'{i}sica - Universidade Federal Fluminense
\\Campus da Praia Vermelha, Av. General Milton Tavares de Souza s/n
\\Gragoat\'{a}, 24.210-346, Niter\'{o}i, Rio de Janeiro, Brasil.}
\email{tsui$@$if.uff.br}
\pagestyle{myheadings}
\baselineskip 18pt
	 
\begin{abstract}

Some critical comments on the prevailing model
of star-disk interaction are made,
in particular, on the rotating nature of the magnetic field lines
and on the application of the magnetohydrodynamic 
frozen-field theorem to the disk plasma.
As an alternative, a unipolar induction model is proposed,
where the magnetic field is stationary in space
and the stellar unipolar electric field $\vec E_{*}$ on the surface
is uploaded to the magnetosphere.
Through the Poynting vector,
the star and the magnetosphere form a coupled system
where the total angular momentum,
consisting the mechanical one of the star,
the electromagnetic one of the magnetosphere,
and the mechanical one of the plasma in the magnetosphere,
is conserved.
The stellar interaction with the accretion disk
is through the projection of the unipolar electric field
$\vec E_{*}$ onto the disk via the equipotential field lines,
generating a disk current and consequently toroidal fields
with opposite signs on both sides of the disk,
with return current loops via the stellar surface.
As a result, magnetic flux is added to the magnetospheric field
in the northern and southern hemispheres
with the disk current sheet as the boundary condition.
This makes the star-disk system an astrophysical site 
where intense magnetic fields are generated
through the rotational energy of the star.
Angular momentum extraction 
from this star-disk system happens 
as the magnetic flux in the magnetosphere
increases to the point 
that exceeds the current carrying capacity of the disk,
leading to a mega scale magnetic eruption
sending Poynting fluxes to space
either isotropically or beamed.

\end{abstract}

\maketitle

\newpage
\section{Rotating Field Line Paradigm}

The star-disk system is a fundamental system in astrophysics
that appears in stellar and galactic scales.
The importance of this system
lies in the enormous angular momentum of the disk
and the extraordinary magnetic field of the magnetic star
that could drive spectacular outflows when adequately combined.
In interactions between a magnetic star and an accretion disk,
the stellar magnetic field lines
are considered firmly anchored on the stellar surface
and corotate with the star
at its angular velocity $\vec\Omega_{*}$
\citep{blandford1976}.
The field lines are like rigid wires,
and plasma kinematics on the field lines
is like 'Beads on a Wire'
\citep{henriksen1971, blandford1982, cao1994, bogovalov1996}.
The interaction of the magnetic field with the plasma on the field line
is described by a rotating field line equation
as in Eq.(2) of \citet{bogovalov1999}
and in Eq.(15) of \citet{tsinganos2000},
so that the field lines could drag the plasma along.

The interaction with an accretion disk on the equatorial plane
with angular velocity $\vec\Omega_{D}$
is governed between the magnetic pressure of the field 
and the ram pressure of the disk plasma.
For a slow rotating star with a weak magnetic field,
the corotation radius $r_{co}$ at which $\Omega_{D}(r)=\Omega_{*}$
is larger than the magnetosphere radius $r_{m}$
(and also the inner boundary of the disk)
which in turn is larger than the Alfv\'{e}n radius $r_{A}$,
where the magnetic pressure balances the plasma ram pressure,
$(r_{A}<r_{m}<r_{co})$,
the field lines will be adverted by the ram pressure
through the magnetohydrodynamic (MHD) frozen-field theorem
to rotate faster in the inner disk $(r_{m}<r<r_{co})$.
An accelerating torque is then imposed to the magnetic star
through the rigid field lines (wires).
Conversely, in the outer disk $(r_{co}<r)$,
the field lines will be advected to rotate slower,
and by the same token a braking torque
is imposed to the star
\citep{lovelace1986, lovelace1995}.
In the case of a fast rotating star with a weak magnetic field,
we have $r_{co}<r_{m}$ and $r_{A}<r_{m}$.
The field lines would be advected by the disk to rotate slower.
As for the case of a fast rotating star with a strong magnetic field
such that $r_{co}$ is located in the magnetosphere $(r_{co}<r_{m}<r_{A})$,
the disk plasma in the interval $r_{m}<r<r_{A}$
will be dragged by the strong field lines to speed up.
The resulting centrifugal force exceeds the gravitational force,
generating the propeller effect
\citep{illarionov1975}
flinging the plasma outward
\citep{romanova2003, ustyngova2006}.
In the outer part of the disk with $r_{A}<r$,
the field lines will be advected by the plasma ram pressure to slow down,
and thus will exert a braking torque on the star.

In this rotating field line paradigm,
there are two basic issues.
The first is the rotating magnetic field line,
which is the primary agent
to tap the rotational energy of the star
and to couple the torque in the star-disk system,
and the second is the anchoring 
of the stellar field lines on the disk
through the frozen-field theorem of a MHD plasma.
Historically, the notion of corotating magnetic field lines
originated from the case of our Sun
\citep{mestel1968},
where the magnetic fields are generated by sources
near the photosphere due to sub-photospheric convections.
These magnetic fields and their field lines corotate with the Sun
because the sources that generate them are corotating with the Sun.
These surface fields are short ranged
and are concentrated in the solar corona.
However, due to the very high corona plasma temperature,
whose heating mechanism is still unsettled,
these field lines are blown into open space by the solar winds
together with the rotating corona plasma,
thus dissipating the angular momentum of the Sun.
This rotating solar wind picture
was then generalized to a body-centered
long range magnetosphere of a magnetic star.
It is assumed that the crust of the stellar surface
firmly anchors the magnetic field lines
such that the stellar surface plays the role of the source surface,
and has become a standard model.
We recall that the Maxwell equations
describe the electromagnetic interactions
via electric and magnetic fields.
Magnetic field line,
defined by the field line equation,
is only a concept devised to visualize
the actions of the magnetic field.
Magnetic field lines, just like velocity stream lines,
do not take physical actions.
For this reason, the field lines should not be treated as rotors.

As for the frozen-field theorem,
it is the most salient feature of a MHD plasma.
According to the MHD plasma model,
which is the one-fluid description of plasma,
the plasma velocity $\vec v$ is described by the equation of motion
where the magnetic force $\vec J\times\vec B$
and the plasma pressure $-\nabla p$ are the driving terms.
This means that the plasma is a magnetized plasma,
and the plasma velocity $\vec v$ represents the center of mass velocity
of the electron and ion plasmas in the two-fluid description of plasma.
The electron equation of motion corresponds to the Ohm law
$\vec E'=\vec E+\vec v\times\vec B=\vec J/\sigma$
with higher order terms neglected,
such as the inertial term and the Hall term, 
where $\vec E'$ is the electric field in the plasma moving frame
and $\sigma$ is the plasma conductivity.
Coupled to the Faraday law of induction, the Ampere law,
the mass continuity equation, and an equation of state,
they form the set of MHD equations.
In the absence of magnetic diffusion with infinite conductivity,
the Faraday law of induction coupled to the Ohm law
determines the response of the magnetic field $\vec B$,
under the action of the plasma flow $\vec v$
(of a magnetized plasma),
to conserve the magnetic flux.
This is known as the frozen-field theorem,
where the compressional and shear Alfv\'{e}n waves
are the notable results of this theorem.

For the interaction of a Kepler disk
with the magnetospheric field of a star,
the disk is basically a gravitational system
whose velocity is set by the stellar gravitational field,
not by the magnetic field of the star.
The plasma of the partially ionized disk
is not a magnetized plasma of the magnetospheric field.
Different from the MHD plasma of the magnetosphere,
the equation of motion of the disk
does not take part in the set of MHD equations,
and the equation of motion of the disk electron plasma
does not correspond to the Ohm law.
Thus, the magnetospheric field and the gravitational disk
do not form a MHD system.
Singling out the Faraday law of induction of the magnetosphere
and the electron plasma equation of motion of the disk
do not warrant the application of the MHD frozen-field theorem. 
Instead, the magnetospheric field and the disk plasma
are two independent systems
that happen to cross each other on the equatorial plane,
and the magnetospheric field 
acts as an external field to the disk plasma.

\newpage
\section{Unipolar Induction Star-Magnetosphere Coupling}

Instead of considering the magnetic field
be anchored on the stellar surface crust,
we consider the magnetosphere as stationary in space
\citep{tsui2015}.
Following \citet{lovelace1995} and \citet{bardou1996},
we regard the star as a perfect conductor
rotating with an angular velocity $\vec\Omega_{*}$
under a magnetic field $\vec B$ in space,
thus generating a unipolar induction electric field $\vec E_{*}$
and a corresponding potential distribution on the stellar surface.
During the built-up of this electric field
and the charge distribution on the stellar surface,
and the subsequent maintenance of the charge distribution in steady state,
there is a current flow in the interior of the star.
This internal current flow generates a braking torque on the star.
Let us take the magnetic moment $\vec m$
and the angular velocity $\vec\Omega_{*}$
both pointing in the upward $\vec z$ axis,
the equatorial region of the star would be charged positively
while the polar region would be otherwise.
With axisymmetry, we can write the magnetic field as

\begin{eqnarray}
\label{eqno1}
\vec B\,=\,A_{0}(\nabla\Psi\times\nabla\phi+\alpha A\nabla\phi)\,
 =\,\vec B_{p}+\vec B_{\phi}\,\,\,,
\end{eqnarray}

\noindent where $A_{0}$ carries the dimension of the poloidal magnetic flux
such that $\Psi$ is a dimensionless poloidal flux function,
$\alpha=A_{\phi}/A_{0}$ represents the relative ratio
of the toroidal flux to the poloidal flux,
and $A$ has the dimension of an inverse scale length
and is related to the axial current through

\begin{eqnarray}
\label{eqno2}
2\pi\alpha A_{0}A\,=\,\mu_{0}I_{z}\,\,\,.
\end{eqnarray}

\noindent According to the field line equation

\begin{eqnarray}
\label{eqno3}
{dr\over B_{r}}\,=\,{rd\theta\over B_{\theta}}\,
=\,{r\sin\theta d\phi\over B_{\phi}}\,\,\,,
\end{eqnarray}

\noindent the poloidal field line
is given by the contour of $\Psi=C$.
Furthermore, the electric field on the stellar surface
can be written as

\begin{eqnarray}
\label{eqno4}
\vec E_{*}\,=\,-V_{*}\nabla\Psi\,=\,-A_{0}\Omega_{*}\nabla\Psi\,\,\,.
\end{eqnarray}

\noindent With the stellar magnetosphere
filled with a quasi-neutral plasma of infinite conductivity
with the Ohm law reading

\begin{eqnarray}
\label{eqno5}
\vec E+\vec v\times\vec B\,
=\,{\vec J\over\sigma}\,
=\,0\,\,\,,
\end{eqnarray}

\noindent the field lines linking the north-south conjugate points
of the star will be equipotential lines,
labeled by the potential distribution on the stellar surface.
Consequently, the unipolar electric field on the stellar surface
can be uploaded to the magnetosphere
which is regarded as stationary in space.
As a result, from Eq.(5), there is a plasma velocity
in the magnetosphere that reads, in spherical coordinates,

\begin{eqnarray}
\label{eqno6}
\vec v_{E}\,
 =\,{1\over B^{2}}(\vec E_{*}\times\vec B_{p})\,
 =\,r\sin\theta\Omega_{*}\hat\phi\,\,\,.
\end{eqnarray}

\noindent This velocity is a rigid rotor velocity
at the angular velocity of the star $\vec\Omega_{*}$
in the toroidal direction.
Although Eq.(6) is simply the solution of Eq.(5),
according to the microscopic description of single particle orbit,
the physical origin of this $\vec v_{E}$
is the Larmor guiding center drift of a magnetized charged particle 
\citep{thompson1962, schmidt1966, jackson1975}.
We remark that $\vec v_{E}$ is not the only plasma drift
in the stationary magnetosphere.
Should we include other transverse forces like plasma pressure,
there would be gradient B and curvature drifts
\citep{tsui2015}.

This unipolar model reproduces the similar results
of the rotating field line approach
as if the plasma were dragged by the rotating field lines.
It offers a dynamic picture of the star
where electric power is continuous generated in the star
and transferred to the magnetosphere 
heating the magnetospheric MHD plasma.
Furthermore, it also removes an internal inconsistency
of the rotating field line picture.
With rotating field lines, the plasma velocity $\vec v$
with respect to the magnetic field $\vec B$
represented by the rotating field lines is null.
Thus, the rotation induced electric field
$\vec E=-\vec v\times\vec B$ should vanish.
However, in the literatures, 
when it comes to calculate the rotation induced electric field $\vec E$,
the magnetic field $\vec B$ is taken as stationary in space instead
to get a nonzero finite $\vec E$,
which amounts to an inconsistency in the rotating field line model.

\newpage
\section{Disk as an External Load}

As the stationary magnetospheric field lines 
intercept the accretion disk on the equatorial plane,
the uploaded electric field is also projected onto the disk plasma.
To describe the star-disk interaction,
considering the disk as a partially ionized plasma,
we follow \citet{bardou1996} to model the disk plasma
as a resistive plasma with current density

\begin{eqnarray}
\label{eqno7}
\vec J\,
 =\,\sigma\vec E^{'}\,
 =\,\sigma(\vec E_{*}+\vec v_{D}\times\vec B_{p})\,
 =\,\sigma(+\Omega_{D}-\Omega_{*})A_{0}\nabla\Psi\,\,\,,
\end{eqnarray}

\noindent where $\vec E^{'}$ is the electric field 
in the disk rotating frame,
$\vec v_{D}$ and $\Omega_{D}$
are the Kepler linear velocity and angular velocity of the disk, 
and $\nabla\Psi$ is pointing inward towards the star.
We note that this equation is not the Ohm law with resistivity
in the MHD system of the magnetosphere.
It is the Ohm law of the disk plasma
under an external magnetic field.
For $r<r_{co}$, the disk current flows inward
generating a braking torque on the disk,
while for $r_{co}<r$, the disk current flows outward
exerting an accelerating torque.
Because of this current flow on the disk,
toroidal magnetic fields with opposite signs
are immediately generated on both sides
of the equatorial disk plane.
Since $\nabla\cdot\vec J=0$,
there are return currents along the disk.
For the inner disk,
these currents follow the equipotential magnetic field lines
to the mid latitudes of the star
and return back to the disk before $r_{co}$,
generating a counter (accelerating) torque on the star.
Likewise is the return currents of the outer disk
which generate a counter (braking) torque
at the high latitude polar region.
Half of the disk current returns from the northern hemisphere,
and the other half returns from the southern part.
In the presence of the toroidal field,
the field line twists around the toroidal direction
as it follows through the poloidal direction,
and Fig.(1) represents the projection
of the twisting field line on the poloidal plane.

Here, by magnetosphere,
we mean the overall closed stellar magnetic field lines
even those intercepting the disk on the equatorial plane,
not just the closed field lines
inside the disk inner boundary $r_{m}$.
Although we have assumed equipotential field lines
such that the unipolar field $\vec E_{*}$
could be projected onto the disk,
this projection is not effective at large distances
either because the magnetospheric plasma density
is not dense enough to ensure perfect conductivity and high plasma mobility,
or the magnetic field is too weak to warrant a MHD state.
Only as the disk accrets closer to the star
that the unipolar field projection becomes effective.
In the rotating field line paradigm,
the toroidal field is generated
by the differential rotation
between the disk plasma and the rotating field lines.
However, there is no clarification of the axial current
to account for the toroidal field.
The toroidal field generation is treated as a magnetostatic problem
\citep{van1994, lovelace1995, lynden1996}.

In the present unipolar model, 
the star-disk current loops
become a site of generating additional magnetic fields,
both toroidal and poloidal,
driven by the rotational energy of the star.
The toroidal fields in the northern and southern hemispheres
have different signs,
which require a self-consistent disk current sheet
as the boundary condition on the equatorial plane.
Integrating the Ampere law 
over the contour of a cross section of the disk 
transverse to a given radial direction,
we get

\begin{eqnarray}
\label{eqno8}
\vec B_{\phi north}\cdot\hat l_{north}
+\vec B_{\phi south}\cdot\hat l_{south}\,
 =\,\mu_{0}J_{r}\Delta z\,
 =\,\mu_{0}\sigma A_{0}(\Omega_{D}-\Omega_{*})
{\partial\Psi\over\partial r}\Delta z\,\,\,,
\end{eqnarray}

\noindent where $\Delta z$ is the thickness of the disk,
$\hat l_{north}$ is the unit vector of the toroidal field 
in the northern hemisphere, likewise is $\hat l_{south}$.
Considering the disk be located outside the corotation radius $r_{co}$,
the disk current is in the $+\hat r$ direction,
the toroidal field in the northern hemisphere 
is in the $-\hat\phi$ direction with $A_{\phi}=-A_{\phi abs}$
where $A_{\phi abs}$ is the absolute value of the magnetic flux.
In the southern side, we have $A_{\phi}=+A_{\phi abs}$.
The boundary condition of the disk current sheet is, therefore,

\begin{eqnarray}
\label{eqno9}
2A_{\phi abs}A(\Psi){1\over r}\,
 =\,-\mu_{0}\sigma(r) A_{0}(\Omega_{*}-\Omega_{D}(r))
{\partial\Psi\over\partial r}\Delta z\,\,\,.
\end{eqnarray}

\newpage
\section{Nonlinear Force-Free Magnetosphere}

 Over the evolution history of a magnetic star,
 we assume that a force-free magnetosphere
 of quasi-neutral plasma
 with the corresponding self-consistent current density is formed,
 described by

\begin{eqnarray}
\label{eqno10}
\vec J\times\vec B=0\,\,\,.
\end{eqnarray}
 
\noindent The toroidal and poloidal components
 of this force-free field equation render

\begin{eqnarray}
\label{eqno11}
A\,=\,A(\Psi)\,\,\,,
\\
\label{eqno12}
{\partial^{2}\Psi\over\partial r^{2}}
 +{1\over r^{2}}
 \sin\theta{\partial\over\partial\theta}
 \left({1\over\sin\theta}{\partial\Psi\over\partial\theta}\right)
 +\alpha^{2}A(\Psi){\partial A(\Psi)\over\partial\Psi}\,
 =\,0\,\,\,.
\end{eqnarray}

\noindent To solve the force-free field,
 we have to specify the source function $A(\Psi)$.
 Should we take a linear dependence for $A(\psi)=a_{0}\Psi$,
 we would get the well known
 linear force-free field solution.
 We prefer to view the magnetosphere as a magnetically stressed system
 and seek a nonlinear description to represent it.
 We thus follow \citet{lynden1994} to consider the coefficient
 as a function of $\Psi$ itself to write

\begin{eqnarray}
\label{eqno13}
A(\Psi)\,=\,a_{0}\Psi^{1/n}\Psi\,
=\,a_{0}\Psi^{1+1/n}\,\,\,,
\end{eqnarray}

\noindent where $a_{0}$ has the dimension
 of an inverse scale length
 and $n$ is the similarity parameter,
 and get

\begin{eqnarray}
\label{eqno14}
r^{2}{\partial^{2}\Psi\over\partial r^{2}}
 +\sin\theta{\partial\over\partial\theta}
 \left({1\over\sin\theta}{\partial\Psi\over\partial\theta}\right)
 +\left(1+{1\over n}\right)
 \alpha^{2}a^{2}_{0}r^{2}\Psi^{1+2/n}\,
 =\,0\,\,\,.
\end{eqnarray}

\noindent Separating the variables
 by writing $\Psi(r,\theta)=R(r)\Theta(\theta)$
 gives
 
\begin{eqnarray}
\label{eqno15}
{1\over R}r^{2}{\partial^{2}R\over\partial r^{2}}\,
 =\,-\left[{1\over\Theta}\sin\theta{\partial\over\partial\theta}
 \left({1\over\sin\theta}{\partial\Theta\over\partial\theta}\right)
 +\left(1+{1\over n}\right)
 \alpha^{2}a^{2}_{0}r^{2}R^{+2/n}\Theta^{+2/n}\right]\,
 =\,n'(n'+1)\,\,\,,
\end{eqnarray}
 
\noindent where $n'$ is the separation constant.
 In terms of the normalized
 radial distance $\xi=r/r_{0}$
 where $r_{0}$ is the radius of the magnetic star,
 the radial function $R(\xi)$ is given by

\begin{eqnarray}
\label{eqno16}
\xi^{2}{\partial^{2}R\over\partial \xi^{2}}\,
 =\,n'(n'+1)R\,\,\,,
\\
\label{eqno17}
R(\xi)\,
 =\,\xi^{n'+1}\,\,\,,\,\,\,\,\,\,\xi^{-n'}\,\,\,.
\end{eqnarray}

\noindent As for the meridian function $\Theta(\theta)$,
 we have

\begin{eqnarray}
\label{eqno18}
\sin\theta{\partial\over\partial\theta}
 \left({1\over\sin\theta}{\partial\Theta\over\partial\theta}\right)
 +n'(n'+1)\Theta\,
 =\,-{(n+1)\over n}
 (\alpha a_{0}r_{0})^{2}\xi^{2}R^{+2/n}\Theta^{1+2/n}
 \,\,\,.
\end{eqnarray}

\noindent With $R(\xi)=\xi^{-n'}$,
 we get the mode equation that reads

\begin{eqnarray}
\label{eqno19}
(1-x^{2}){d^{2}\Theta(x)\over dx^{2}}
 +n'(n'+1)\Theta(x)\,
 =\,-{(n'+1)\over n'}
 (\alpha a_{0}r_{0})^{2}\Theta^{1+2/n'}(x)
 \,\,\,,
\end{eqnarray}

\noindent where $x=\cos\theta$ and we have taken $n'=n$.
 From the field line equation,
 the twist angle of the field line in space $\phi(x)$
 and the corresponding toroidal magnetic field
 are governed by

\begin{eqnarray}
\label{eqno20}
d\phi(x)\,
 =\,-(\alpha a_{0}r_{0}){1\over n'}
 {1\over (1-x^{2})}\Theta^{1/n'}(x)dx
 \,\,\,,
\\
\label{eqno21}
B_{\phi}\,
 =\,{A_{0}\over r_{0}^{2}}
 (\alpha a_{0}r_{0}){1\over\xi\sin\theta}
 \Psi^{1+1/n'}
 \,\,\,.
\end{eqnarray}

\noindent Once $\Theta(x)$ is solved from the mode equation,
 the twist angle $\phi(x)$ can be integrated
 to get its profile in space.
 
\newpage
\section{Stressed Magnetosphere}

With the right side of Eq.(19) null for $\alpha=0$ or $A_{\phi}=0$,
 the linear solution is

\begin{eqnarray}
\label{eqno22}
\Theta(x)\,
 =\,(1-x^{2}){dP_{n'}(x)\over dx}\,\,\,,
\end{eqnarray}

\noindent with integer $n'=1,2,3,...$
 where $P_{n'}(x)$ is the Legendre polynomial.
 With $n'=n=1$, the poloidal field lines are given by
 
\begin{eqnarray}
\label{eqno23}
\Psi(\xi,\theta)\,=\,R(\xi)\Theta(x)\,
 =\,{\sin^{2}\theta\over\xi}\,=\,C\,\,\,,
\end{eqnarray}

\noindent which corresponds to the dipole field.
 For $\alpha\neq 0$, the meridian function $\Theta(x)$ is stressed
 through the right side of Eq.(19)
 generating a nonlinear $\Theta(x)$ 
 with $n'$ deviating from integers.
 To represent the magnetospheric field as an external field,
 not anchored on and twisted by the disk plasma,
 we take the boundary condition of $\Theta(x)$
 on the disk at $x=0$ as
 
\begin{eqnarray}
\label{eqno24}
{d\Theta(x)\over dx}\,
 =\,0\,\,\,,
\end{eqnarray}

\noindent in order to assure
 the poloidal field line be normal to the conducting disk.
 Solving for the mode equation
 with $R(\xi)=\xi^{-n'}$ and with $n'=n$,
 the response of the eigenvalue $\alpha a_{0}r_{0}$
 to the separation constant $n'$ is shown in Fig.2.
 The separation constant $n'$
 begins from $n'=1$ for $\alpha a_{0}r_{0}=0$,
 and as $n'$ decreases the eigenvalue increases.
 However, as the separation constant
 goes beyond $n'=0.720$,
 there is no eigenvalue $\alpha a_{0}r_{0}$
 to allow an eigenfunction $\Theta(x)$.
 In this sense, $n'=0.720$ is the stress limit
 of the nonlinear description of $\Theta(x)$.
 The stressed eigenfunctions $\Theta(x)$
 whose nonlinearity changes with $n'$
 are shown in Fig.3.
 Three values of $n'=0.998, 0.853, 0.720$ are chosen
 corresponding to $\alpha a_{0}r_{0}=0.10, 0.50, 0.59$
 respectively. 
 Furthermore, the poloidal field line contours
 with $C=0.2$ are plotted in Fig.4.
 The contour $C=0$ is the polar axis.
 With $C=0.2$, the field lines of Fig.4
 are originated in the polar region,
 and they are the stressed counterpart
 of the linear dipole field lines.
 As the contour value $C$ increases,
 the poloidal field lines will come
 from a region of lower latitude of the star
 and will reach the inner part of the disk
 closer to the star.
 The plasma disk, which is not shown in this figure,
 lies on the equatorial plane
 extending from $\xi>\xi_{co}$ outward.
 The twist angles in space $\phi(x)$
 are illustrated in Fig.5,
 and the dimensionless toroidal fields $B_{\phi}(x)$,
 without the dimensional $A_{0}/r_{0}^{2}$ factor,
 at a given radius $\xi=5$ on the disk plane at $x=0$
 are illustrated in Fig.6.
 In Fig.5, the twist angle in space
 increases with the eigenvalue,
 in particular on the disk plane.
 In Fig.6, the toroidal field on the disk
 increases with the eigenvalue,
 thus with the twist angle.
 Within this stressed magnetosphere model,
 the field line twisting
 reaches its limit as $n=n'=0.720$
 corresponding to $\alpha a_{0}r_{0}=0.59$.
 The toroidal magnetic field saturates at this limit.
 The dimensionless toroidal fields $B_{\phi}(x)$,
 without the dimensional $A_{0}/r_{0}^{2}$ factor,
 along the trajectories of three poloidal field lines
 are indicated in Fig.7.
 As a result, Fig.4 and Fig.7 together
 show spiralling magnetospheric field lines
 down to the star.

\newpage
\section{Angular Momentum Extraction}

Let us now examine the current sheet boundary condition
on the equatorial plane.
First of all, the circumference of the disk increases as $r$,
and to maintain a current flow on the disk,
the plasma conductivity has to scale as $1/r$ by axisymmetry.
In terms of the normalized radial distance $\xi$,
$\sigma(r)=\kappa/r=\kappa/(r_{0}\xi)$.
Furthermore, we consider the inner disk boundary 
be located at a fair distance beyond $\xi_{co}$
such that $(\Omega_{*}-\Omega_{D}(\xi))$ 
can be regarded as insensitive to $\xi$.
This point can also be supported 
by the accelerating torque on the disk
which reduces the angular velocity gradient across the disk.
With these considerations,
and using Eq.(13) and $R(\xi)=\xi^{-n'}$,
the $\xi$ dependence has canceled out 
from both sides of the boundary condition, Eq.(9),
leaving only

\begin{eqnarray}
\label{eqno25}
2A_{\phi abs}a_{0}r_{0}\Theta^{1/n}(x)\,
 =\,\mu_{0}\kappa A_{0}n(\Omega_{*}-\Omega_{D})\Delta z\,\,\,.
\end{eqnarray}

\noindent This condition contains the disk parameters
$\sigma, \Omega_{D}, \Delta z$,
the field parameters $A_{\phi abs}, A_{0}, a_{0}$,
and the meridian function $\Theta(x)$ evaluated at $x=0$.
As the field parameters evolve over time,
the disk parameters have to evolve accordingly.
To bind the opposing extraordinary toroidal fields together,
the disk should carry an ultra high current density
which would certainly heat up the disk to extreme temperatures.
This enhances the degree of ionization of the disk plasma
which allows the disk to carry more current.
Furthermore, the corresponding blackbody radiations
of the disk temperature
might be related to quiescent X-ray emissions.
As the toroidal field in the magnetosphere builds up,
the magnetic pressure compresses the disk increasing its density,
thus further enhancing the plasma conductivity
and allowing the disk to carry even more current.
The compression of the disk by the magnetic pressure
counteracts the propeller effect
caused by the accelerating torque on the disk
and keeps the disk on the equatorial plane.
Under this scenario,
we can therefore regard the magnetosphere as a site 
where the rotational energy of the star
is stored in the form of magnetic energy
through the toroidal and poloidal fields
via the unipolar induction.

By uploading the potential distribution on the stellar surface
to the equipotential field lines of the magnetosphere,
the unipolar stellar electric field $\vec E_{*}$,
generated by the angular momentum of the star, 
is transferred to the magnetosphere.
Consequently, there is an electromagnetic momentum in the magnetosphere 
given by the Poynting vector

\begin{eqnarray}
\label{eqno26}
\mu_{0}\vec S\,
=\,\vec E_{*}\times\vec B\,
=\,V_{*}A_{0}(\nabla\Psi\cdot\nabla\Psi)\nabla\phi
\mp V_{*}A_{\phi abs}A\nabla\Psi\times\nabla\phi
\,\,\,.
\end{eqnarray}

\noindent The first term is the toroidal momentum
whereas the second term is the poloidal momentum
with the upper/lower sign for the southern/northern hemisphere.
Summing over the entire magnetosphere,
the southern and the northern poloidal momentum cancel each other,
leaving only the toroidal momentum.
The corresponding electromagnetic angular momentum
with respect to the center of the star is
$\vec L=\vec r\times\vec S$.
This angular momentum has a toroidal part and a poloidal part.
Summing over the entire magnetosphere,
it is clear that the toroidal part vanishes
while the poloidal part adds up to a z component along the polar axis
as is the angular momentum of the star.
Most of the time, we often think of electromagnetic fields as fields only.
But the momentum and angular momentum
associated to the electromagnetic fields
are just as real and physical as the mechanical counterparts.
These electromagnetic momentum and angular momentum
in the stationary magnetosphere
can further be coupled to the magnetospheric plasma 
through the $\vec E\times\vec B$ drift,
as if it were dragged by the rotating field lines.
We therefore have a star-magnetosphere system
where angular momentum of the star
can be coupled to the magnetosphere
and further to the plasma there.
On the other hand, the star losses angular momentum
due to the unipolar current flow in the interior of the star,
which exercises a braking torque on the star.
This star-magnetosphere system underlines the conservative nature 
of the total angular momentum, mechanical and electromagnetic.

We remark that in this model
the star-disk system acts like an astrophysical power circuit
where the star is the generator and the disk is the external load.
Through the current flows between the star and the disk,
the toroidal and poloidal force-free magnetic fields are enhanced
in the northern and southern hemispheres.
This provides a site where extraordinary toroidal fields are generated 
at the expense of the stellar rotational energy.
The toroidal flux stored in the magnetosphere
is directly related to the current carrying capacity of the disk,
which is a function of the plasma conductivity.
As the magnetic flux keeps increasing,
it eventually exceeds the current carrying limit of the disk,
generating a violent magnetic eruption
and sending Poynting fluxes into space
either isotropically or beamed.
With the electromagnetic angular momentum being extracted,
it leaves the star to establish anew its magnetosphere
with the disk as a plasma source
to refill the nearby stellar space through diffusion.

\newpage
\section{Conclusions}

We have pointed out in the current rotating field line model
of star-disk interaction two basic problems.
First, magnetic field line, like velocity stream line,
is a concept devised to help visualize
the actions of the magnetic field.
Magnetic field line is not a real physical variable
and it cannot take actions to drag plasmas to corotate.
This generates an internal inconsistency
in this rotating field line model
where the rotation induced electric field should be zero
because there is no relative velocity
between the rotating plasma and the rotating field.
However, the rotation induced electric field
has always been calculated
as if the magnetic field were stationary in space.
Second, the disk is a gravitational system
and it does not form an ideal MHD system
with the stellar magnetosphere.
The two systems are not described
under the same set of MHD equations.
Consequently, the frozen-field theorem cannot be applied.
We have, therefore, proposed a unipolar induction
star-disk interaction model
with stationary magnetic fields in space
where the unipolar stellar electric field
is uploaded to the magnetosphere
generating a star-magnetosphere coupled system
with angular momentum conservation,
the mechanical one of the star
and the electromagnetic one of the magnetosphere.
The electromagnetic one of the magnetosphere
can further be coupled to the plasma there
through the $\vec E\times\vec B$ drift,
as if the plasma were dragged by the rotating field lines.
In this model, the interaction between the star and the disk
is through the projection of the uploaded
stellar electric field onto the disk,
thus generating a current flow on the disk
with return current loops via the stellar surface,
forming a site
where intense toroidal and poloidal magnetic fields
can be generated from the rotational energy of the star.
The opposing toroidal magnetic fields
in the northern and southern hemispheres
are held together by the disk current sheet
under the self-consistent boundary condition,
which is a function of the disk parameters.
Angular momentum of this star-disk system can be extracted
when the magnetic flux in the magnetosphere grows so much 
that it exceeds the current carrying capacity of the disk
leading to the repulsion of the mega toroidal fields
and sending Poynting fluxes into space.

\acknowledgments

\appendix

\clearpage
\begin{figure}
\plotone{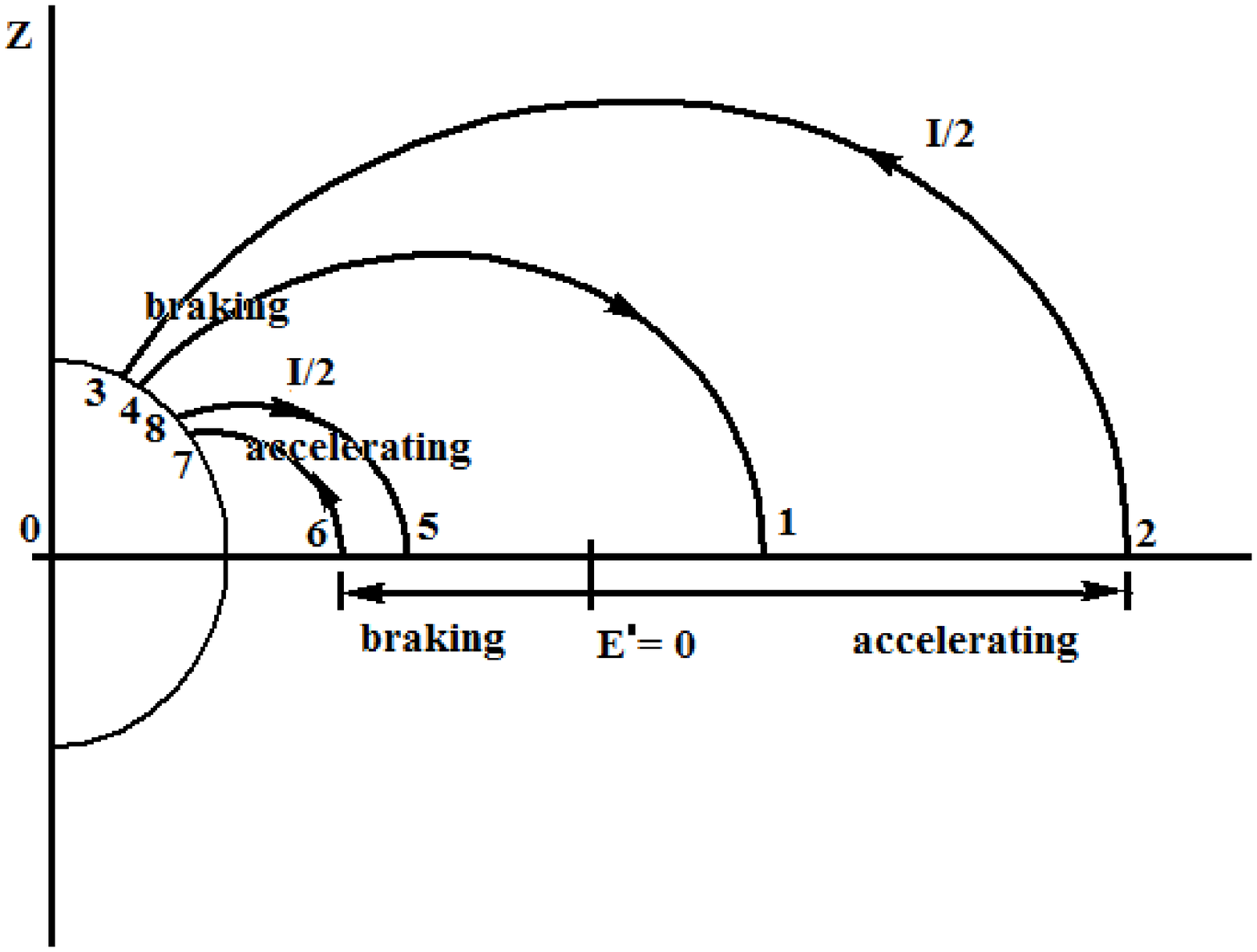}
\caption{This figure illustrates the unipolar electric field interaction scheme.
The position $\vec E'=0$ indicates the null of the electric field 
in the rotating frame of the disk plasma,
from which the current of the resistive disk
flows inward towards the star and outward away from the star.
The inner current loop is completed through the equipotential field lines
to the stellar surface and back (5-6-7-8-5),
likewise is the outer loop (1-2-3-4-1).
Through these current loops, torque and counter torque
are applied on the disk and star respectively, and vice verse,
and corresponding strong toroidal magnetic fields can also be generated.}
\label{fig.1}
\end{figure}

\clearpage
\begin{figure}
\plotone{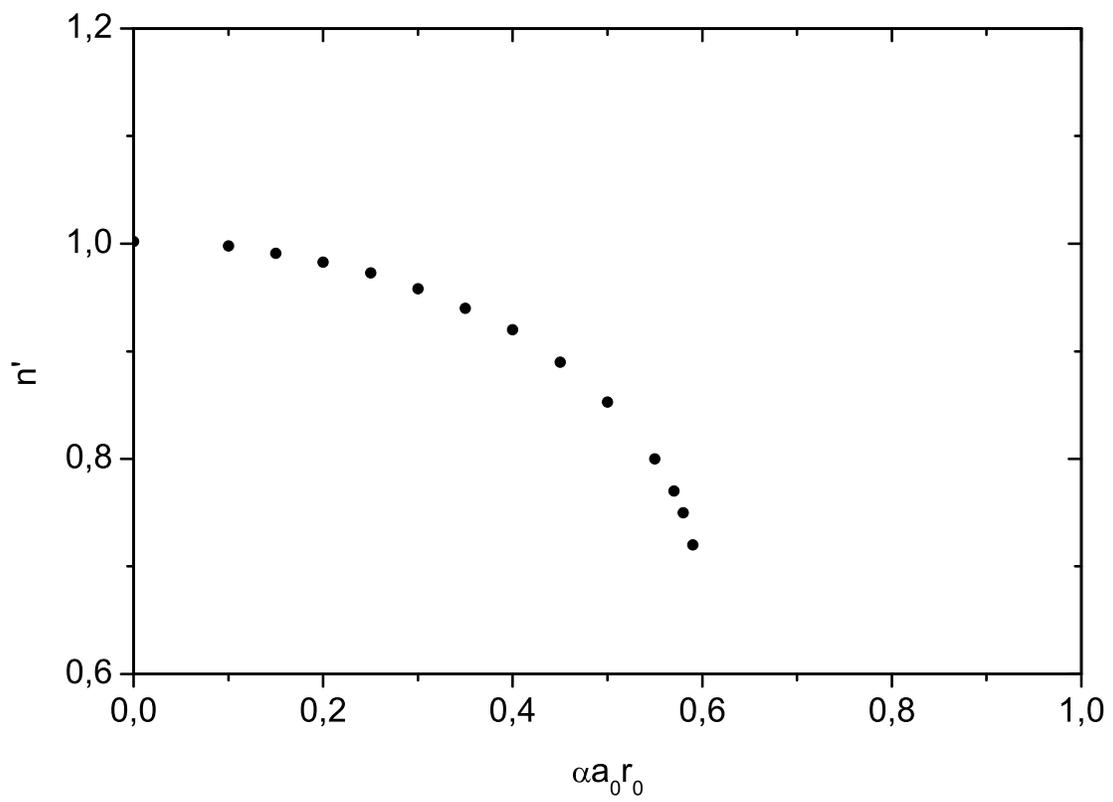}
\caption{The eigenvalue $\alpha a_{0}r_{0}$ of the mode equation
 is shown as a function of the separation constant $n'$.}
\label{fig.2}
\end{figure}

\clearpage
\begin{figure}
\plotone{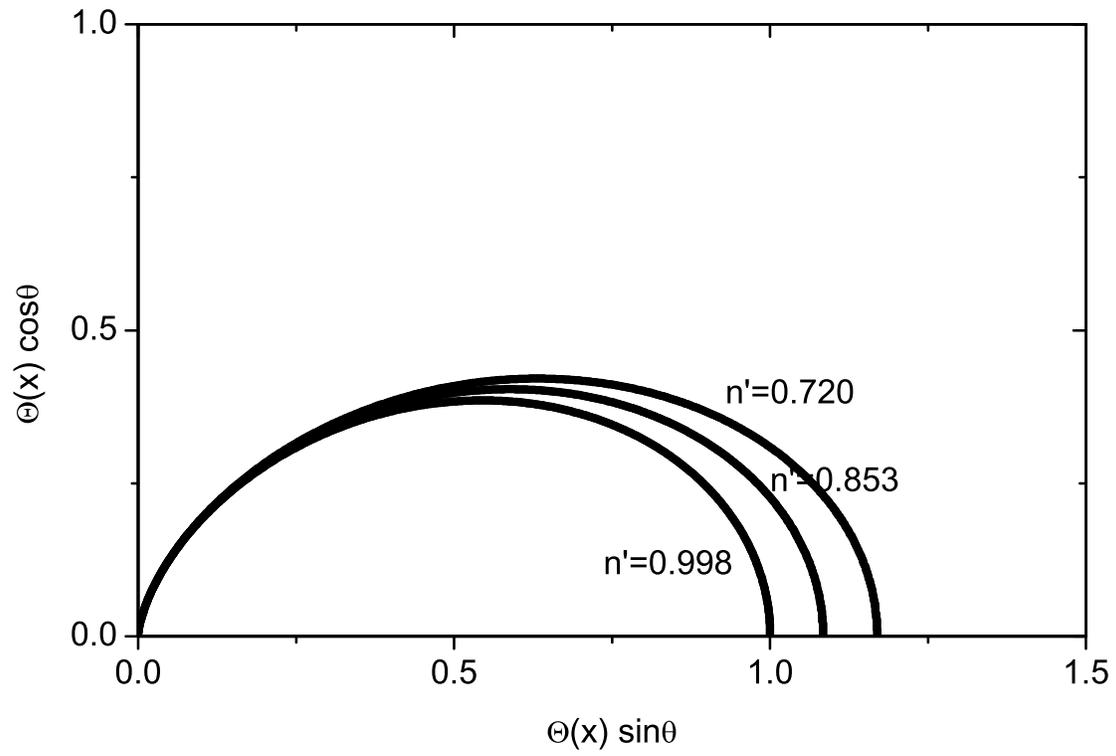}
\caption{The stressed eigenfunctions $\Theta(x)$
 are shown in a polar plot
 for $\alpha a_{0}r_{0}=0.10, 0.50, 0.59$
 corresponding to $n'=0.998, 0.853, 0.720$ respectively
 for the three plots from left to right.}
\label{fig.3}
\end{figure}

\clearpage
\begin{figure}
\plotone{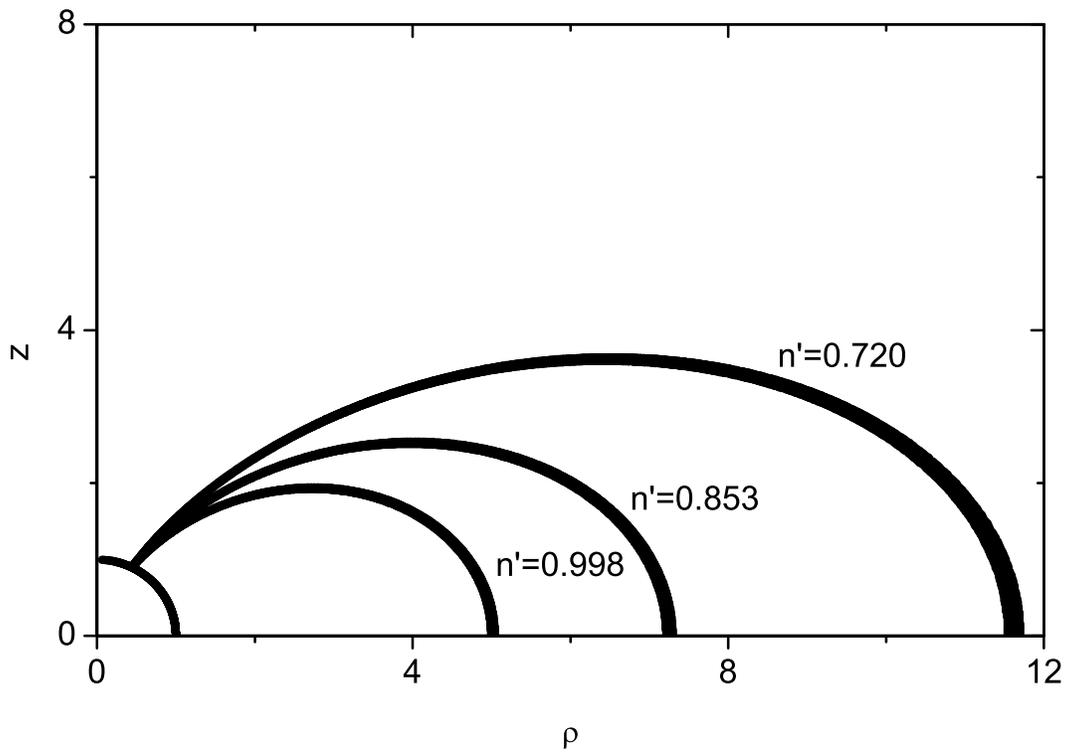}
\caption{The poloidal field lines $\Psi(\xi,x)=C$ with $C=0.20$
 are shown in a cartesian $(\rho,z)$ plot,
 where $\rho=\xi\sin\theta$,
 in units of normalized distance $\xi$
 for the three eigenfunctions $\Theta(x)$
 of Fig.2 from left to right.
 The contour of $C=0$ is the polar axis.
 As the contour value $C$ increases,
 the contours move closer to the equatorial plane
 and to the star.}
\label{fig.4}
\end{figure}

\clearpage
\begin{figure}
\plotone{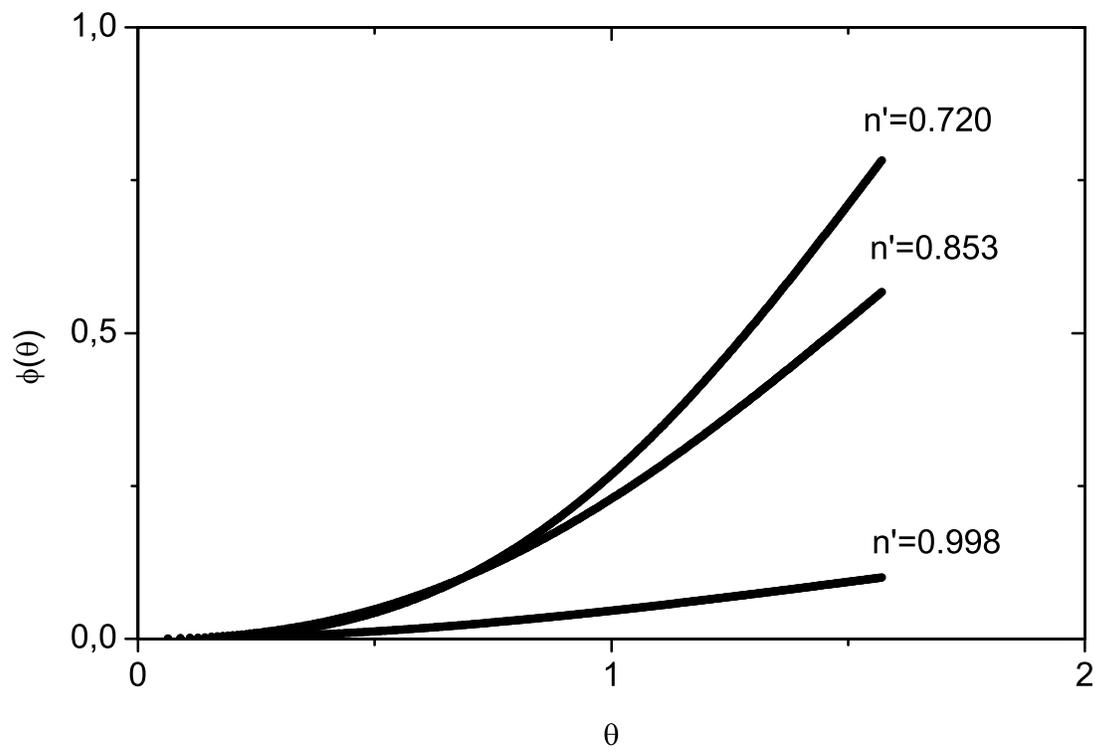}
\caption{The twist angles $\phi(\theta)$ in space
 are plotted against the polar angle $\theta$
 for the three field lines $\Psi(\xi,x)=C=0.20$
 with $\alpha a_{0}r_{0}=0.10, 0.50, 0.59$
 for the lower, middle, upper curves respectively.}
\label{fig.5}
\end{figure}

\clearpage
\begin{figure}
\plotone{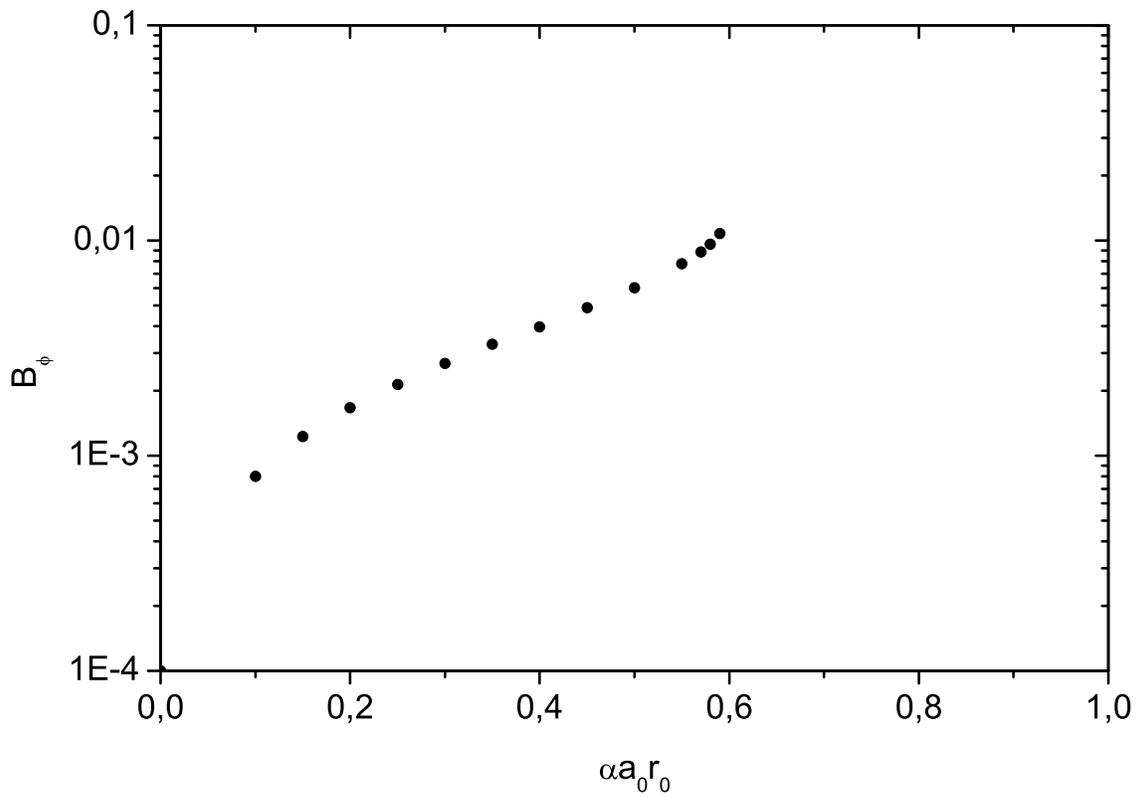}
\caption{The dimensionless toroidal magnetic fields $B_{\phi}$
 on the disk plane at $\xi=5$
 are plotted against the eigenvalues $\alpha a_{0}r_{0}$.}
\label{fig.6}
\end{figure}

\clearpage
\begin{figure}
\plotone{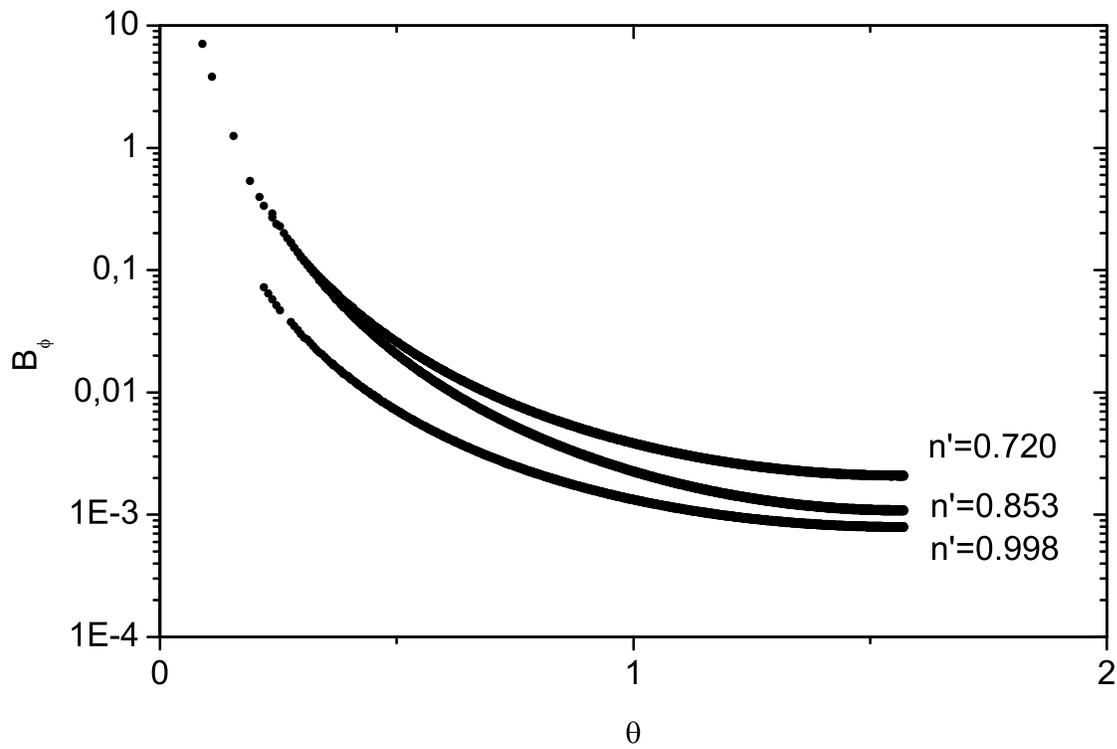}
\caption{The dimensionless toroidal magnetic fields $B_{\phi}$
 are plotted along the three poloidal field lines.}
\label{fig.7}
\end{figure}

\end{document}